\documentstyle[psfig]{mn}

\newif\ifAMStwofonts

\def\simgt{\,\hbox{\rlap{\raise 0.425ex\hbox{$>$}}\lower 0.65ex\hbox{$\sim$}}\,}
\def\simlt{\,\hbox{\rlap{\raise 0.425ex\hbox{$<$}}\lower 0.65ex\hbox{$\sim$}}\,}
\def\arcsec{^{\prime\prime}}
\def\nid{\noindent}

\input{psfig}

\ifoldfss
  \ifCUPmtlplainloaded \else
    \NewTextAlphabet{textbfit} {cmbxti10} {}
    \NewTextAlphabet{textbfss} {cmssbx10} {}
    \NewMathAlphabet{mathbfit} {cmbxti10} {} 
    \NewMathAlphabet{mathbfss} {cmssbx10} {} 
  \fi
  \ifAMStwofonts
    \ifCUPmtlplainloaded \else
      \NewSymbolFont{upmath} {eurm10}
      \NewSymbolFont{AMSa} {msam10}
      \NewMathSymbol{\upi}     {0}{upmath}{19}
      \NewMathSymbol{\umu}     {0}{upmath}{16}
      \NewMathSymbol{\upartial}{0}{upmath}{40}
      \NewMathSymbol{\leqslant}{3}{AMSa}{36}
      \NewMathSymbol{\geqslant}{3}{AMSa}{3E}

    \fi
  \fi
\fi 

\ifnfssone
  \newmathalphabet{\mathit}
  \addtoversion{normal}{\mathit}{cmr}{m}{it}
  \addtoversion{bold}{\mathit}{cmr}{bx}{it}
  \newmathalphabet{\mathbfit} 
  \addtoversion{normal}{\mathbfit}{cmr}{bx}{it}
  \addtoversion{bold}{\mathbfit}{cmr}{bx}{it}
  \newmathalphabet{\mathbfss} 
  \addtoversion{normal}{\mathbfss}{cmss}{bx}{n}
  \addtoversion{bold}{\mathbfss}{cmss}{bx}{n}
  \ifAMStwofonts
    \ifCUPmtlplainloaded \else
      \UseAMStwoboldmath
      \makeatletter
      \new@mathgroup\upmath@group
      \define@mathgroup\mv@normal\upmath@group{eur}{m}{n}
      \define@mathgroup\mv@bold\upmath@group{eur}{b}{n}
      \edef\UPM{\hexnumber\upmath@group}
      \new@mathgroup\amsa@group
      \define@mathgroup\mv@normal\amsa@group{msa}{m}{n}
      \define@mathgroup\mv@bold\amsa@group{msa}{m}{n}
      \edef\AMSa{\hexnumber\amsa@group}
      \makeatother
      \mathchardef\upi="0\UPM19
      \mathchardef\umu="0\UPM16
      \mathchardef\upartial="0\UPM40
      \mathchardef\leqslant="3\AMSa36
      \mathchardef\geqslant="3\AMSa3E
    \fi
  \fi
\fi 

\ifnfsstwo
  \DeclareMathAlphabet{\mathbfit}{OT1}{cmr}{bx}{it}
  \SetMathAlphabet\mathbfit{bold}{OT1}{cmr}{bx}{it}
  \DeclareMathAlphabet{\mathbfss}{OT1}{cmss}{bx}{n}
  \SetMathAlphabet\mathbfss{bold}{OT1}{cmss}{bx}{n}
  \ifAMStwofonts
    \ifCUPmtlplainloaded \else
      \DeclareSymbolFont{UPM}{U}{eur}{m}{n}
      \SetSymbolFont{UPM}{bold}{U}{eur}{b}{n}
      \DeclareSymbolFont{AMSa}{U}{msa}{m}{n}
      \DeclareMathSymbol{\upi}{0}{UPM}{"19}
      \DeclareMathSymbol{\umu}{0}{UPM}{"16}
      \DeclareMathSymbol{\upartial}{0}{UPM}{"40}
      \DeclareMathSymbol{\leqslant}{3}{AMSa}{"36}
      \DeclareMathSymbol{\geqslant}{3}{AMSa}{"3E}
    \fi
  \fi
\fi 

\ifCUPmtlplainloaded \else
  \ifAMStwofonts \else 
    \def\upi{\pi}
    \def\umu{\mu}
    \def\upartial{\partial}
  \fi
\fi

\title[Image Separation of Lensed QSOs]
{Image Separation vs. Redshift of Lensed QSOs: 
Implications for Galaxy Mass Profiles}
\author[L. L. R. Williams]
       {Liliya L. R. Williams\thanks{Email: {\bf \tt llrw@ast.cam.ac.uk}} \\
Institute of Astronomy, Madingley Road, Cambridge, CB3 0HA }
\date{Accepted for publication in MNRAS}
\pubyear{1997}

\begin{document}

\maketitle

\newcommand{\fmmm}[1]{\mbox{$#1$}}
\newcommand{\scnd}{\mbox{\fmmm{''}\hskip-0.3em .}}
\newcommand{\scnp}{\mbox{\fmmm{''}}}

\begin{abstract}
Recently, Park and Gott reported an interesting observation: image 
separation of lensed QSOs declines with QSO redshift more precipitously 
than expected in any realistic world model, if the lenses are taken to 
be either singular isothermal spheres or point masses. In this Letter 
I propose that the observed trend arises naturally if the lensing 
galaxies have logarithmic surface mass density profiles that gradually 
change with radius. If the observed lack of central (odd) images is 
also taken into account, the data favor a universal dark 
matter density profile over an isothermal sphere with a core. 
Since the trend of image separation vs. source redshift is mostly 
a reflection of galaxy properties, it cannot be straightforwardly 
used as a test of cosmological models. Furthermore, the current 
upper limits on the cosmological constant may have to be 
revised.
\end{abstract}

\begin{keywords}
gravitational lensing -- cosmology -- galaxies: structure
\end{keywords}

\section{Introduction}\label{intro}

Since gravitational lensing takes place over cosmological distances, 
it can in principle be used to measure cosmological parameters, such
as mass density, $\Omega$, and cosmological constant, $\Lambda$. 
For example, the frequency of multiply imaged QSOs is a sensitive 
function of $\Lambda$, and therefore the observed abundance of lensed 
QSOs has been used to place upper limits on $\Lambda$ (Turner 1990;
Fukugita et al. 1992; Kochanek 1996). Another example is 
the relation between image separation, $\theta_{im}$ and source 
redshift, $z_s$ of multiply split QSOs, which mostly depends 
on the sum of $\Omega$ and $\Lambda$, and thus should serve as a 
good indicator of the curvature of the Universe. 
(Gott, Park \& Lee 1989; Fukugita et al. 1992, Park \& Gott 1997,
hereafter PG97). If the Universe is flat and galaxies are singular 
isothermal spheres, the $\theta_{im}$--$z_s$ relation should be 
flat, i.e. independent of source 
redshift. If the Universe is open, there should be a mild decline
of image separations with redshift; a $\sim 20\%$ decline is
expected between $z_s\sim 1.5$ and 5, in an extreme $\Omega=0$ case.

The mildness of the expected $\theta_{im}$--$z_s$ trend can be 
explained as follows. For any given lens, and fixed lens and source 
redshifts the image separation of images with comparable brightness 
is quite independent of source impact parameter. Since optimal 
lens redshift for any high redshift source is roughly the same, 
$z_l\sim 0.5$, and angular diameter distances vary little past 
$z\sim 0.5$, the observed image separation should change little with 
$z_s$. If a population of lenses is considered, having a range
of properties, like scale lengths and velocity dispersions,
there will be a corresponding spread in image separations, but
still little or no trend with $z_s$.

Yet the observed $\theta_{im}$ of lensed QSOs declines strongly
with $z_s$ (See Figure 1 of PG97, or Figure~\ref{relation} of 
this paper.) PG97 present a list of source redshifts and image 
separations of 20 multiply imaged QSOs. The source redshifts span a 
range between 1.4 and 4.5, and image separation between 0.5$\arcsec$ 
and 7$\arcsec$. There is a drop of at least a factor of 4 in image 
separation from the low-$z_s$ to high-$z_s$ cases. According to PG97,
in a flat Universe such distribution can be ruled out at 99\% 
confidence level, while in the most extreme empty Universe,
it is ruled out at 97\%. To make the confidence level decrease 
below 95\%, one would need to `remove' 2--3 largest separation or 
highest redshift cases from the sample. (See PG97 for a 
discussion of statistical significance, possible errors, etc.) 

In this Letter, I suggest that the trend, which is due to the lack 
of high redshift wide separation lensed QSOs is naturally reproduced 
if a set of three conditions is met, (i) galaxies' central mass profiles 
have logarithmic slopes that change with radius, (ii) there is a 
dispersion in galaxy properties, like central surface mass densities 
or velocity dispersion, and (iii) the characteristic length scale of 
galaxy dark matter halos varies with galaxy luminosity as 
$r \propto L^a$, where $0<a\simlt 0.5$.

\section{Model}\label{model}

In this section I describe the lensing model, and introduce some
simplifying assumptions about the lenses, galaxies, and sources, 
QSOs. I do not aim to produce a detailed model because the small 
number of lensed QSOs in the current sample does not necessitate it.
I need not assume any particular cosmological model, because the 
effect implied by the present model is considerably stronger than 
that of cosmology. 

\subsection{Model Assumptions}

\subsubsection{Lenses}

I consider two types of galaxy mass profile, an isothermal
sphere with a core, ISC, and a universal dark matter profile, NFW,
derived from numerical simulations of Navarro, Frenk \& White 
(1995, 1996). These profiles were chosen because their logarithmic
surface mass density gradually flattens with decreasing radius. 
The reason for using these, as opposed to constant power law 
profiles, like a singular isothermal sphere, is discussed in 
Section~\ref{implications}. Both ISC and NFW provide a plausible 
description of the dark matter halos of galaxies. ISC is commonly 
used because of the observed shapes of the rotation curves of 
spiral galaxies. Additionally, based on the results of the HST 
Snapshot Survey, Maoz \& Rix (1993, hereafter MR93) conclude that 
ellipticals must possess ISC-type dark matter halos in order to 
reproduce the observed image separation distribution of multiply 
imaged QSOs. The NFW profile has been recently claimed as a universal 
dark matter halo profile describing halos of gravitationally bound 
structures over 4 orders of magnitude in mass. It is also supported 
on theoretical grounds (Evans \& Collett 1997). 
I assume that all galaxy-lenses are circularly symmetric.

A description of the lensing properties of these profiles 
can be found in Schneider et al. (1992), Bartelmann (1996), and 
Williams \& Lewis (1997). It suffices to say here that ISC is roughly 
flat within core radius $r_c$, and has an isothermal density profile 
outside. The vertical normalization of the profile is fixed by its 
central surface mass density, $\kappa_0$, in terms of critical
density for lensing, $\Sigma_{crit}$. NFW is singular at the centre,
its logarithmic slope gradually steepens from $\rho\propto r^{-1}$ 
at the centre, to $\rho\propto r^{-3}$ at large radii, and is
isothermal around the characteristic scale length, $r_s$. The vertical 
normalization of its projected surface mass density is proportional 
to $\kappa_s$, which is given in terms of $\Sigma_{crit}$. 

As was pointed out by PG97, the scatter in the observed 
$\theta_{im}$--$z_s$ points is large. One of the main sources of 
scatter is probably the spread in galaxy-lens properties.
For example, a singular isothermal
sphere lens of a given $\sigma_v$ produces a constant bending 
angle at the lens, and hence image separation varies as
$D_{ls}/D_{os}$, where the $D$'s are the lens--source and 
observer--source angular diameter distances. If all lensing 
galaxies had the same $\sigma_v$, one would expect the observed 
angular image separations to be proportional to the corresponding 
$D_{ls}/D_{os}$ values, whereas a plot of $D_{ls}/D_{os}$ vs.
$\theta_{im}$, in 8 systems where both $z_s$ and $z_l$ are known 
reveals no correlation at all, and a scatter in $\theta_{im}$ of 
a factor of a few larger than the scatter in $D_{ls}/D_{os}$. 
This strongly suggests that the spread in galaxy-lens properties 
is important. Therefore I assume a {\it family} of galaxy-lenses, 
each having the same scale length, $r_c$ and $r_s$ for ISC and 
NFW respectively, but a range of $\kappa_0$ and $\kappa_s$,
which is equivalent to a range in velocity dispersions.

I assume that $r_c$ and $r_s$ and galaxy luminosity function
do not evolve with redshift, at least in 
the interval most relevant here, i.e. optimal lens redshift 
for high-$z$ sources, $z_l\sim 0.3-1$. This assumption is supported
by the recent spectroscopic observations of Lilly et al. (1995), 
which indicate that the luminosity function of elliptical galaxies,
i.e. the galaxy population that is believed to provide the bulk of 
the lensing optical depth (see MR93), evolves very little from 
$z\sim$1 to the present.

When the source impact parameter is sufficiently small, all the
NFW lenses, and ISC lenses with $\kappa_0>1$ produce 3 images, 
whereas the observed multiply imaged systems mostly have 
an even number of images, either 2 or 4. This is thought to be 
because the central (odd) image is demagnified below visibility. 
I will return to the importance  of the central odd image later, 
in Section~\ref{results}. The circularly symmetric lenses considered 
here cannot produce 5 image systems; observed 4(+1) cases are a 
property of elliptical lenses. However the present treatment will 
not suffer if only symmetric lenses are considered.

For sources at large redshifts, the optimal lens redshift increases
very slowly with $z_s$ in all cosmologies except for the most extreme
$\Lambda$ dominated cases. So I will assume that $z_l$ is constant,
independent of $z_s$. The critical surface mass density for lensing, 
$\Sigma_{crit}$ changes little with $z_s$, if $z_l$ is fixed, and 
sources are at high $z_s$. For example, in an $\Omega$=0.3 open
Universe, moving the source from $z_s$=1 to 5 reduces $\Sigma_{crit}$ 
by 17\%, if $z_l=0.2$. Therefore, I will assume that galaxy-lens 
projected surface mass density is given by $\kappa_0$ and $\kappa_s$, 
with no dependence on redshifts. 

\subsubsection{Sources}

Analytic fits to QSO luminosity function (QLF) usually take the form 
of a double power law, with steep bright-end slope and shallow 
faint-end slope; the transition occurs at some characteristic 
luminosity, $L_0$, which evolves with redshift, and is commonly 
parameterized as $L_0(z_s)\propto (1+z_s)^{\alpha}$. 
Boyle et al. (1990) derive a
pure luminosity evolution for $z\simlt 2$ QSOs, with
$\alpha$=3.2, while Hewett et al. (1993) conclude that QLF changes
shape with redshift, and that evolution must slow down beyond 
$z\sim 1.5$ compared to the predictions of Boyle et al. For 
$2\simlt z\simlt 3$ they derive $\alpha\sim$1.5. QSO evolution at
higher redshifts, up to 5 is less constrained, though it is sometimes
assumed that the shape of the QLF does not evolve beyond $z\sim 2$,
while $L_0$ evolves such that it `slides' brightward along the high
luminosity part of the QLF (Wallington \& Narayan 1993). 
I assume that at any given redshift there are no QSOs 
brighter than $L_0$, and that QLF at $L<L_0$ is 
$N_{QSO}(L)\propto L^{-s}$, with $s=1.2$ (MR93). 
I adopt the $\alpha$ parameterization of QSO evolution for 
high redshifts, and derive results for $\alpha$=0, and 2. 

\subsection{Results}\label{results}

We are interested in two lensing properties of the galaxy-lenses: 
image separation and total image magnification. Given a galaxy-lens,
a source can have a range of impact parameters, each resulting in
a different image magnification, and image separation, even though
the latter tends to stay rather constant, roughly equal to twice the
Einstein ring radius of the lens. To account for a range of source 
impact parameters, I calculate the impact parameter weighted averages
of image magnification, $\langle\mu\rangle$ and image separation, 
$\langle\theta\rangle$ as a function of $\kappa_0$ or $\kappa_s$, 
for ISC and NFW respectively.

\begin{figure}
\vbox{
\centerline{
\psfig{figure=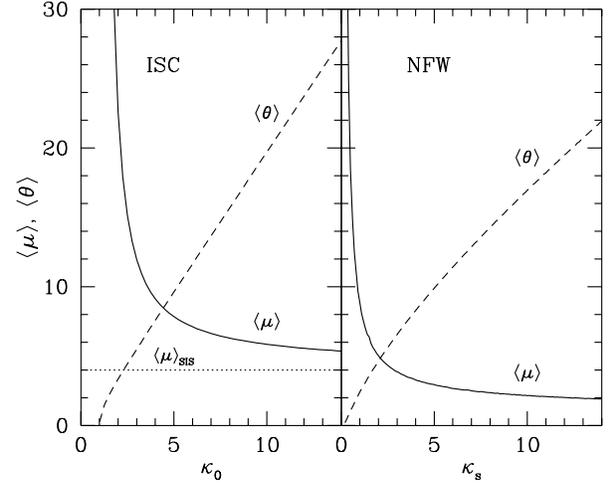,width=3.25in,angle=0}
}
\caption[]{Magnification and image separation vs. the central
surface density of two different types of lenses: ISC and NFW.
Both magnification and image separation are impact parameter weighted
averages of these quantities. The dotted line is the impact parameter 
weighted average magnification of a singular isothermal sphere.
Note that for ISC and NFW profiles the magnification and image
separation go in the opposite sense with increasing $\kappa_0$
and $\kappa_s$, while for SIS the average magnification is always
a constant (see Section~\ref{implications}).
\label{ISCNFW2}
}
}
\end{figure}

\begin{figure}
\vbox{
\centerline{
\psfig{figure=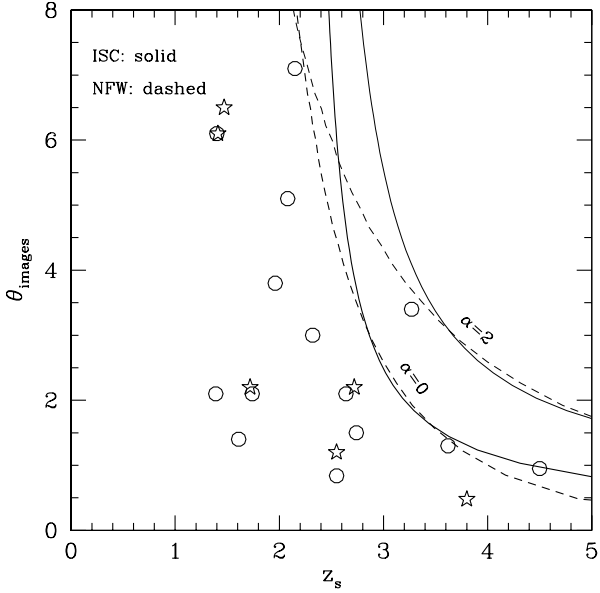,width=3.25in,angle=0}
}
\caption[]{Image separation vs. source redshift. The points are
the observed multiply imaged QSOs (taken from Table 1 of PG97); 
stars represent the lenses from the HST Snapshot Survey (MR93),
circles are from other, ground based surveys. Predictions of the
isothermal sphere with a core, ISC and universal dark matter halo,
NFW models (labeled) were normalized vertically and horizontally,
as described in Section~\ref{results}. Both ISC and NFW profiles 
can account for the observed trend.
\label{relation}
}
}
\end{figure}

Figure~\ref{ISCNFW2} shows the results. Vertical normalization of 
image separation is irrelevant for now. The important feature
is that for both profiles the image separation increases with 
$\kappa_0$ or $\kappa_s$, while total image magnification decreases. 
In other words, given a population of galaxy-lenses with a range
of $\kappa_0$ or $\kappa_s$, image magnification goes inversely with
image separation. Now, image magnification couples to source redshift, 
as follows. Flux of an $L_0$ QSO is given by, 
\vskip 0.07in
${\hskip 0.39in}f\propto \mu(z_s) L_0(z_s)/{D_L(z_s)}^2$\\
${\hskip 0.72in} \propto \mu(z_s) (1+z_s)^\alpha/{D_A(z_s)}^2(1+z_s)^4$\\
${\hskip 0.72in} \propto \mu(z_s) (1+z_s)^{\alpha-4},{\hskip 1.3in}(1)$
\vskip 0.07in
\nid where $\mu(z_s)$ is total image magnification, and $D_L$ and
$D_A$ are the luminosity and angular diameter distances respectively.
In the second and third lines I used the fact that 
$D_L = (1+z)^2 D_A \propto (1+z)^2$, since angular diameter 
distance stays roughly constant for high-$z_s$ sources.
So to make it into a magnitude limited sample, with a certain
$f_{lim}$, an $L_0$ QSO at $z_s$ has to be magnified by at least 
$$\mu(z_s)\propto f_{lim} (1+z_s)^{4-\alpha},\eqno(2)$$ 
with fainter QSOs magnified more. Because larger redshift QSOs 
undergo larger magnifications, a fixed $f_{lim}$ of any given survey 
translates into a lower limit on $\langle\mu\rangle$ as a function 
of $z_s$, which in turn implies an upper 
limit on $\langle\theta\rangle$, as read off from Figure~\ref{ISCNFW2}. 
The resulting relation is plotted as solid (ISC) and dashed (NFW)
curves in Figure~\ref{relation}, for two values of QSO luminosity
function evolution parameters, $\alpha$; 0, and 2. Both sets of 
curves reproduce the upper envelope of the observed points quite 
well. The paucity of observed points does not allow to differentiate 
between $\alpha$ of 0 and 2, or between ISC and NFW profiles.

The curves are scaled vertically and horizontally to roughly 
match the observed distribution. Adjusting the scaling in the vertical
direction yields the angular size of $r_c$ and $r_s$ respectively.
For ISC angular core radius is 0.37$\arcsec$, while for NFW
angular scale radius is 1.25$\arcsec$. For typical lens redshifts
of 0.5 these translate into $r_c=1.4 h^{-1}$ kpc, and
$r_s=4.6 h^{-1}$ kpc. These values are roughly what one would derive 
from modelling individual multiple-image systems using ISC/NFW profiles. 
Scaling the curves in the horizontal direction gives the average 
{\it minimum} magnification 
of lensed QSOs as a function of $z_s$. For $\alpha=0$ models,
$\langle\mu\rangle/f_{lim} = 0.039 (1+z_s)^4$, and for $\alpha=2$,
$\langle\mu\rangle/f_{lim} = 0.4 (1+z_s)^2$; i.e. if QSO characteristic 
luminosity does not evolve, a lower limit on magnification of QSOs
at $z=3$ is $\langle\mu\rangle/f_{lim}$=10, whereas it is 6.4 if 
$L_0(z_s)\propto (1+z_s)^2$. Surveys with brighter $f_{lim}$
have higher $\langle\mu\rangle$, but the dependence on $(1+z_s)$ 
is weak, and most existing surveys have $f_{lim}$ of 1--2 magnitudes 
within each other. 

Let us summarize the conclusions thus far. Roughly, 
Figure~\ref{ISCNFW2} implies that average image magnification is 
inversely proportional to the galaxy central surface mass density,
$\langle\mu\rangle\propto\kappa_{0,s}^{-1}$; and that average image
separation is directly proportional to the galaxy central surface mass 
density,
$\langle\theta\rangle\propto\kappa_{0,s}$. In general, the observed 
image separation is also proportional to the characteristic scale 
length of the galaxy-lens, so 
$\langle\theta\rangle\propto\kappa_{0,s} r_{c,s}$, 
but since I have so far assumed that all galaxies have the same 
characteristic scale length, I have not used the 
latter dependence. The scale length can be a function of central 
concentration, $r_c\propto\kappa_{0,s}^\beta$. Combining these
relations gives,
$\langle\mu\rangle\propto\langle\theta\rangle^{-{1\over 1+\beta}}$,
and together with eq.(2) it implies,
$$\langle\theta\rangle\propto (1+z_s)^{-(1+\beta)(4-\alpha)}\eqno(3).$$
With $\beta=0$ this equation basically
reproduces the curves in Figure~\ref{relation}. Note that $\beta$
does not have to be 0, any value between roughly -0.5 and 2 will do,
i.e. the total exponent in eq.~(3) should be around --{\it few}.
I will return to the discussion of $\beta$ and its implications for
the dark matter halos of galaxies in Section~\ref{discussion}.

Thus both ISC and NFW galaxy-lens models can reproduce the observed 
$\theta_{im}$--$z_s$trend, and neither is preferred based on these 
observations alone. However, another widely documented observation 
can break the tie; all multiply imaged QSOs lack the central odd-numbered 
image, i.e. observed cases are either doubles or quadruples. This has 
long been interpreted as evidence in favor of centrally condensed 
galaxy profiles, which would demagnify the central image below visibility.
The other possible explanation, dust obscuration in the centres of 
galaxies, is largely ruled out because radio observations of multiply 
split QSOs also do not reveal central images (Myers et al. 1995).
I apply this argument here: ISC model predicts that $\kappa_0\sim 1-2$ 
models should have central images of brightnesses comparable to the 
primary image, while with the NFW models the central images of low 
$\kappa_s$ lenses should be 5-10 times demagnified compared to the 
primary image. This rules strongly in favor of the NFW model. 

\section {Implications for galaxy mass profiles}\label{implications}

The key feature of the galaxy lens population that allows to reproduce
the envelope is that image separation goes inversely with image 
magnification. This allows magnification bias, which acts through
image magnification, to couple image separation to source redshift.
This property of a lens population is a direct consequence of the 
individual galaxies having a changing logarithmic slope, as in
the ISC and NFW models. Lenses that do not have this property,
for example singular isothermal spheres and point masses,
cannot reproduce the observed envelope.

For a singular isothermal sphere (SIS) the average impact parameter 
weighted magnification can be calculated to be 4, independent of 
the velocity dispersion, $\sigma_v$, while image separation
is proportional to $\sigma_v^2$. So $\langle\mu\rangle$, being a
`universal' 
constant, does not couple to $\langle\theta\rangle$, and thus SIS
cannot reproduce the observed trend. The property that
$\langle\mu\rangle$ is constant is not unique to SIS, but can be
generalized to any family of mass profiles with no in-built scale
length, like single power-laws and point masses. In such models
the only parameter is the absolute mass normalization of the profile,
which determines the size of Einstein ring radius, and hence the
image separation. However, $\langle\mu\rangle$ in such self-similar
profiles is a constant, independent of absolute mass normalization. 

Therefore galaxies must have changing logarithmic surface mass density 
profiles in order to be able to reproduce the $\theta_{im}$--$z_s$
relation.

\section{Discussion}\label{discussion}

There are two interesting aspects in the observed distribution of
image separations vs. source redshift. In this Letter I set out 
to explain one of these, the lack of high redshift wide separation
lenses, i.e. the $\theta_{im}$--$z_s$ anticorrelation.
It turns out that the model presented in Section~\ref{model} also 
naturally explains the
other interesting feature of the $\theta_{im}$--$z_s$ plot,
namely the existence of wide separation cases; there
are 8 lensed QSOs with $\theta_{im}>3\arcsec$. Most models
currently found in the literature have trouble predicting a large 
population of wide separation lenses; for example
MR93 predict the peak in $\theta_{im}$ distribution at 
$\simlt 1\arcsec$, with vanishingly small number of cases above 
$3\arcsec$. So it has been argued that large $\theta_{im}$ cases 
are not due to isolated galaxies, but are the result of 
cluster-aided galaxy lensing. QSO 0957+561, with 6.1$\arcsec$ 
between its two images, is adduced as supporting evidence. 
However, if cluster-aided galaxy lensing is the correct explanation,
then wide separation lenses should be found at all redshifts,
and not just at low $z_s$, as is currently the case 
(Figure~\ref{relation}).
Furthermore, PG97 show that with the help of cluster lensing,
image separation should {\it increase} with $z_s$. Since this
is clearly not observed, cluster aided lensing is probably not 
important in general (0957 must be a special case), and the 
large number of wide separation lenses needs an explanation.

Let us derive the distribution of image separations as predicted 
by the model of Section~\ref{model}. 
(I will use the ISC model because all the
derivations required in this section can be carried out analytically
with this model. However the general arguments presented here will 
also apply to the NFW profile.) So far only the dark matter 
halo properties have been discussed. These need to be related to
the observable galaxy properties, like luminosity. I will adopt 
relations similar to those used in MR93 (their Section 2.2.1);
$$r_c\propto L^a,{\hskip 0.12in} M/L \propto L^b,{\hskip 0.07in} 
{\rm hence} {\hskip 0.12in} \kappa_0\propto L^{1-2a+b}\eqno(4).$$ 
Optical effective radii of ellipticals are observed to scale as 
$L^{1.2}$ (Lauer 1985),
and since it is commonly assumed that they are linearly proportional 
to the galaxies' dark matter scale length, $a$ is usually taken to 
be 1.2. From the observations of the fundamental plane of ellipticals
(Kormendy \& Djorgovski 1989), $b\approx 0.25$. Similar assumptions 
were made by Kochanek (1996). Parameter $\beta$ of eq.~(3) is
$\beta=a/(1-2a+b)$\footnote{Note that MR93 and Kochanek (1996) 
assumed $a$=1.2, and $b=0.25$, and hence effectively their 
$\beta=-1.04$, which would not reproduce the $\theta_{im}$--$z_s$ 
anticorrelation, see eq.~(3).}. 

The frequency of multiply imaged QSOs as a function of image
separation is given by,
$${dF\over d\langle\theta\rangle}={dN(L)\over dL}\,[r_c y_r(L)]^2\,
{dL\over d\langle\theta(L)\rangle}\,
\Bigl({\langle\mu(L)\rangle\over f_{lim}}\Bigr)^s, \eqno(5)$$
where $dN(L)/dL \propto L^{-1.2} e^{-L}$ is the Schechter luminosity 
function (Schechter 1976); 
$L$ is in units of $L_*$, and I assume a constant slope 
of -1.2. The lensing cross section for a 
galaxy of luminosity $L$ is given by the radial caustic in the 
source plane, whose radius is $y_r(L)=(\kappa_0^{2/3}-1)^{3/2}$.
The image separation is approximately equal to the diameter of the
tangential critical curve in the lens plane, 
$\langle\theta\rangle\approx 2 r_c \sqrt{\kappa_0^2-1}$. 
Using eq.~(4), 
$${d\langle\theta(L)\rangle\over dL}
={2 r_{c,*}L^{a-1}(\kappa_{0}^2[1-a+b]-a)\over
(\kappa_{0}^2-1)}.\eqno(6)$$
Here, $r_{c,*}$ and $\kappa_{0,*}$ refer to the core radius and
$\kappa_0$ of an $L_*$ galaxy. I take $r_{c,*}$ to be 1.4$h^{-1}$kpc,
equal to the constant core radius derived in Section~\ref{results}.
The value of $\kappa_{0,*}$ is then calculated from $r_{c,*}$
and an assumed asymptotic line of sight velocity dispersion,
$\sigma_*=300$km~s$^{-1}$; $\kappa_{0,*}=5$. The last term in
eq.~(5) is the magnification bias.

\begin{figure}
\vbox{
\centerline{
\psfig{figure=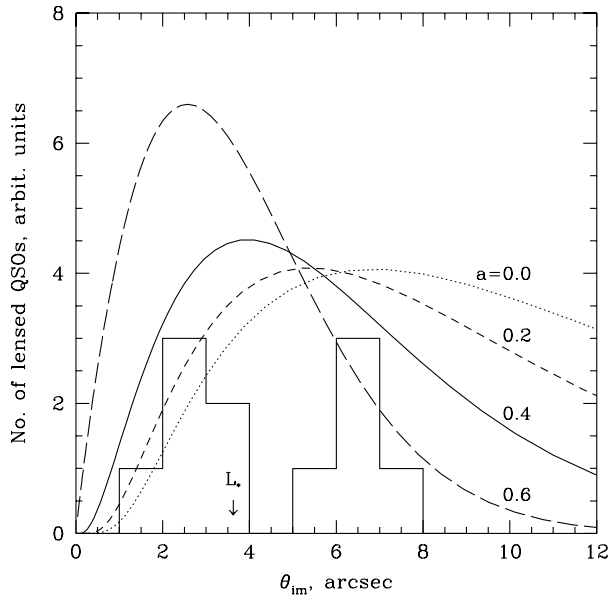,width=3.25in,angle=0}
}
\caption[]{Predicted image separation distribution for lensed QSOs.
The curves are labeled by parameter $a$, defined by $r_c\propto L^a$.
Vertical normalization is arbitrary. The predictions are valid for
$z_s$ where the magnification bias effect, discussed in 
Section~\ref{results} and shown in Figure~\ref{relation}, does not 
impose an upper limit on $\theta_{im}$ of high-$z_s$ lensed QSOs, i.e. 
for $z_s\simlt 2.5$. Accordingly, the observed distribution to which
the curves apply, is restricted to $z_s<2.5$ and is shown as the solid
histogram. Values of $a\sim 0.4$ seem to describe the observed
distribution adequately. In these models image separation is a
monotonically increasing function of galaxy luminosity; image 
separation produced by an $L_*$ galaxy is indicated by an arrow.
\label{xsection}
}
}
\end{figure}

The function $dF/d\langle\theta\rangle$ of eq.~(5) is plotted in 
Figure~\ref{xsection}, for a range of $a$ corresponding to the 
allowable range of $\beta$, in eq.~(3), $a$=0.0, 0.2, 0.4, and 0.6;
$b=0.25$ was used throughout. The angular splitting by an $L_*$ 
galaxy is denoted by an arrow. These predicted distributions apply 
to the source redshift range where the upper envelope, described 
by eq.~(3) and plotted in Figure~\ref{relation} does not cut in, 
i.e. for $z_s\simlt 2.5$. The solid histogram is the observed 
distribution for the same source redshift cutoff. Even though 
the small number of currently observed lenses
does not allow to make any precise conclusions, it is apparent that 
$a\sim 0.4$ reproduces the histogram quite well. The corresponding
value of $\beta$ is 0.9, which is perfectly consistent with the
allowed range of $\beta$, -0.5 to 2 (see eq.~[3]). 
{\it Thus, $a\sim 0.4$ ($\beta\sim 0.9$) reproduces both the 
$\theta_{im}$--$z_s$ anticorrelation, and the observed
frequency of wide separation lenses.} In this scenario, 
the $\theta_{im}\sim 7\arcsec$ lenses are due to $\sim 2 L_*$
galaxies, with a mass within
10$h^{-1}$kpc of $3.7~10^{11}h^{-1}M_\odot$. 

The major deviation of the present model from those found in the
literature is in the value of $a$, the power law index relating
the dark matter characteristic scale length of a galaxy to its
luminosity. As mentioned earlier, $a$ is usually assumed to be 1.2,
while consistency with lensing observations in the framework of
the present model implies $a\sim0.4$.

\section{Conclusions} 

If the observed distribution of image separation vs. source
redshift of lensed QSOs is real and not a result of
false lenses, like physical QSO pairs, then it has the following
implications for galaxy dark matter halos: 
(i) galaxies must have changing logarithmic surface mass density 
profiles, with a `universal dark matter' model being preferred over an
isothermal sphere with a core, (ii) there must be a spread in galaxy 
properties, like the central surface mass density, (iii) the 
characteristic length scale of dark matter halos should scale with 
luminosity as $L^a$, where $a\sim 0.4$.

The corollary of the present model is that higher $z_s$ 
multiply imaged QSOs are statistically more magnified, and are  
preferentially lensed by galaxies with lower central surface mass 
densities and intrinsic luminosities. 
These predictions can be tested if each lens case is modelled 
individually, which will become possible when the lensing galaxies 
of most multiply split QSOs are detected.

The strong dependence of the observed $\theta_{im}$--$z_s$ 
relation on galaxy properties means that it cannot be easily used 
as a test for the curvature of the Universe unless we know a lot 
more about the mass distribution of galaxies. Furthermore, because 
of the implications the present model has on the lensing properties 
of galaxy population, like the dependence of the predicted frequency
of lensed QSOs on parameter $a$ (see Figure~\ref{xsection}), 
the current upper limits on $\Lambda$ may have to be revised.

\section*{Acknowledgments}

I am grateful to Joachim Wambsganss for useful suggestions.
I would like to acknowledge the support of PPARC
fellowship at the Institute of Astronomy, Cambridge, UK.

\end{document}